\documentclass[aps,prl,twocolumn,showpacs,superscriptaddress,floatfix]{revtex4}
\usepackage{graphicx,color}
\begin{document}

  \author{S.\ Miyahara}
  \affiliation{
    Multiferroics Project (MF), ERATO,
    Japan Science and Technology Agency (JST), 
    c/o Department of Applied Physics, The University of Tokyo, 
    7-3-1 Hongo, Tokyo 113-8656, Japan}
  \author{N.\ Furukawa}
  \affiliation{
    Multiferroics Project (MF), ERATO,
    Japan Science and Technology Agency (JST), 
    c/o Department of Applied Physics, The University of Tokyo, 
    7-3-1 Hongo, Tokyo 113-8656, Japan}
  \affiliation{
    Department of Physics and Mathematics, 
    Aoyama Gakuin University, 
    5-10-1 Fuchinobe, Sagamihara, Kanagawa 229-8558, Japan}

  \title{
    Theory of electric field induced one-magnon resonance 
    in cycloidal spin magnets}
  \date{\today}

  \begin{abstract}
    We propose a new mechanism to induce a novel one-magnon 
    excitation by the electric component of light
    in cycloidal spin states, {\it i.e.} so called electromagnon process.
    We calculated optical spectra in the 
    cycloidal spin structures
    as observed  in multiferroic perovskite manganites $R$MnO$_3$
    where novel magnetic excitations induced by 
    oscillating electric fields are observed.
    When symmetric spin-dependent electric polarizations are introduced,
    we have light absorptions at terahertz frequencies
    with one- and two-magnon excitations driven 
    by the electric component of light.
    Our results show that some parts of optical spectra
    observed experimentally at terahertz frequencies  
    are one-magnon excitation absorptions.
  \end{abstract}


  \maketitle 

  Multifferoics, where ferromagnetism and 
  ferroelectricity coexist, have attracted
  both experimental and 
  theoretical interests~\cite{tokura06,eerenstein06,cheong07}.
  Recently, ferroelectric perovskite manganites
  $R$MnO$_3$ ($R = $ Tb, Dy, Gd, and others) 
  are especially paid attention to 
  due to colossal magnetoelectric 
  effects~\cite{kimura07},
  since ferroelectricities are strongly 
  associated with their cycloidal spin orders, and as a result,
  they have strong magnetoelectric couplings.
  Theoretically, such a strong coupling between the ferroelectricity and 
  the cycloidal spin state can be understood 
  by considering a spin-current induced polarization
  described by $\vec{P}_A \propto \vec{e}_{ij} \times
  (\vec{S}_i \times \vec{S}_j)$~\cite{katsura05,mostovoy06,sergienko06}.
  Once the cycloidal spin state is realized, 
  an electric polarization breaks out
  whose direction is perpendicular to
  the propagation and the helicity vectors of the
  cycloidal spin structure.
  One of the experimental evidences to confirm the spin-current scenario
  is that, in $R$MnO$_3$, the direction of polarizations
  can be controlled by changing the direction 
  of the cycloidal plane with external magnetic 
  fields~\cite{kimura07}.
  For example, in TbMnO$_3$ the polarization flops from $P \| c$ 
  to $P \| a$ by 
  a magnetic field induced spin flop  from $bc$ to $ab$ cycloidal states.

  One of the hot topics in these multifferoics materials $R$MnO$_3$
  is electric field induced spin excitations, or {\em electromagnon},
  observed in an optical spectroscopy at terahertz 
  frequencies~\cite{pimenov06,pimenov06b,aguilar07,pimenov07,
    kida08,kida08b,takahashi08}.
  In the pioneering work by Pimenov {\it et al.},
  they observe magnetic absorptions
  in ${\rm TbMnO_3}$ and ${\rm GdMnO_3}$ for $E \| a$~\cite{pimenov06}. 
  Especially, absorption around $2$ meV in ${\rm TbMnO_3}$
  is expected to be strongly related to ferroelectricity
  induced in the spin-current model.
  In fact, the spin-current model predicts
  that the electric component of light
  perpendicular to the cycloidal spin plane,
  {\it i.e.} $E \| a$ in ${\rm TbMnO_3}$, 
  can excite a one-magnon excitation~\cite{katsura07}.
  Thus, excitations observed in ${\rm TbMnO_3}$
  were considered as such one-magnon excitations.

  However, further detailed investigations revealed that terahertz absorptions 
  in $R{\rm MnO_3}$ could not simply be understood by 
  the theories proposed so far.
  Main unclear feature is concern with 
  a selection rule in the spiral ordered phase,
  which  does not depend on $R$ site ions, temperature, 
  external magnetic fields, and hence directions of spins.
  This is in contradiction with
  the prediction of the spin-current model $E \perp P$.
  For example, at the lowest
  temperature region of the spiral spin state in zero magnetic field,
  $P \| c$ in ${\rm TbMnO_3}$, and ${\rm DyMnO_3}$  
  whereas $P \| a$  in 
  ${\rm Eu_{1-x}Y_{x}MnO_3}$ and ${\rm Gd_{0.7}Tb_{0.3}MnO_3}$
  are observed, and then
  $E\| a$ and $E\| c $ absorptions are expected, respectively.
  However, in reality, 
  the strongest absorptions are always observed
  for $E \| a$~\cite{pimenov06,pimenov06b,
    aguilar07,pimenov07,kida08,kida08b,takahashi08}.
  Moreover, recent experiments in ${\rm DyMnO_3}$
  confirm that,
  even when the polarization flops from $P \| c$ to $P \| a$ 
  by the external magnetic fields,
  the selection rule hardly changes~\cite{kida08}.

  In this paper, we propose a mechanism of absorption
  originated from low-energy spin excitations induced by 
  electric fields at terahertz frequencies
  based on a theory proposed for far infrared 
  absorptions in antiferromagnets~\cite{tanabe65,moriya66,moriya68},
  {\it i.e.}, magnon absorptions due to virtual electron hoppings 
  are taken into account while
  we neglect the phonon assisted magnon absorption.
  We assume that antiferroelectric spin-dependent polarization 
  proportional to symmetric spin term $S_i \cdot S_j$ 
  is dominant. As observed in N\'{e}el ordered state,
  the conventional simultaneous two-magnon absorption
  is possible even in cycloidal spin state.
  However, our results indicate that novel 
  {\it one-magnon} absorption
  induced by the electric component of light is 
  dominant in most of the cycloidal spin states. 
  Because of the crystal structure,
  only $E \| a$  component produces the one-magnon absorption
  in $R{\rm MnO_3}$. 
  We can reproduce some parts of experimental observation by
  considering a one-magnon absorption quite well.

\begin{figure}
    \begin{center}
      \includegraphics[width=0.65\columnwidth]{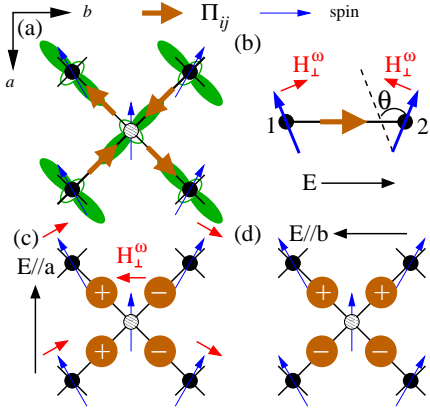}
    \end{center}
    \caption{(Color online) 
      (a) Cycloidal spin structures, orbital ordering pattern 
      on Mn site, and direction of 
      symmetric spin dependent polarizations.
      (b) Non collinear spin structure on a dimer.
      Applying an electric field along the bond
      produces an effective transverse staggered fields $H^\omega_\perp$.
      (c) For $E \| a$, signs of $(\vec{E}^{\omega} \cdot \vec{\Pi})$ 
      are shown on each bond. Spins are oscillated by
      an effective transverse fields shown by (red) arrows,
      which induces one-magnon resonance due to electric fields.
      (d) For $E \| b$, signs of $(\vec{E}^{\omega} \cdot \vec{\Pi})$ 
      are shown. The effective fields cancel out.
    }
    \label{fig:one-magnon}
  \end{figure}

  The spin Hamiltonian under the external electric field
  can be written as $H = H_0 - \vec{E} \cdot \vec{P}$,
  where $H_0$ is a spin Hamiltonian like a Heisenberg Hamiltonian,
  and $\vec{P}$ is an electric dipole moment which depends 
  on the spin configurations. 
  In $R$MnO$_3$, the electric dipole 
  moment $\vec{P}_{ij}$ associated
  with a pair of spins on the nearest-neighbor bonds
  have been calculated by a microscopic theory 
  for the Mn-O-Mn $180^\circ$ bond~\cite{jia07}.
  In spin dependent polarizations, the symmetric spin dependent term 
  \begin{equation}
    \vec{P}_{S} 
    = \sum_{n.n.} \vec{\Pi}_{ij} \, (\vec{S}_i \cdot \vec{S}_j)
    \label{eq:polarization_S}
  \end{equation}
  is likely dominant~\cite{moriya68,jia07}.
  Since such a kind of polarization is realized due to 
  the $3x^2-r^2/3y^2-r^2$ orbital ordering at Mn sites,
  the directions of $\vec{\Pi}$ are determined
  as shown in Fig.~\ref{fig:one-magnon} (a). 
  For the cycloidal spin structure,
  the local symmetric spin polarization
  terms $\vec{P}_{S}$ are aligned antiferroelectrically
  (see Fig.~\ref{fig:one-magnon} (a)).
  This additional term does not violate the fact that the total
  ferroelectricity occurs solely due to the antisymmetric 
  spin dependent term
  $P_{A} \propto \vec{S}_i \times \vec{S}_j$.

  We can easily see that one-magnon can be induced 
  by the electric component of light for noncollinear 
  spin structures. 
  Let us concentrate on one bond connecting site 1 and 2
  (Fig.~\ref{fig:one-magnon} (b)).
  When we apply the oscillating electric field $\vec{E}^{\omega}$
  along the bond, Hamiltonian of the field
  is $- \vec{E}^{\omega} \cdot \vec{P}_s = 
  - (\vec{E}^{\omega} \cdot \vec{\Pi}) \vec{S}_1 \cdot \vec{S}_2$.
  Since $\vec{S}_1$ and $\vec{S}_2$ are noncollinear,
  the spin $\vec{S}_1$ is oscillated by an effective field 
  perpendicular to $\vec{S}_1$,
  $H_{1 \perp}^\omega =  (E^\omega \cdot \Pi) S \sin \theta$.
  The same calculation for the spin $\vec{S}_2$ indicates that
  the effective oscillating transverse fields are staggered:
  $H_{1 \perp}^\omega = - H_{2 \perp}^\omega$.
  Thus, electric component of light can induce 
  a one-magnon resonance caused by effective staggered 
  transverse fields,
  once a noncollinear ground state is realized. 

  The symmetric spin dependent polarization 
  gives us the selection rule $E \| a$ straight forwardly.
  In $R$MnO$_3$, the propagation vector of cycloidal state 
  is aligned to $b$ directions, {\it i.e.},
  spins along $a$ direction are uniform.
  When we apply the electric field $E \| a$,
  $\vec{E}^{\omega} \cdot \vec{\Pi}$ is uniform 
  (staggered) along $a$ ($b$) direction.
  As a result, electric fields couple with the spin structures.
  Thus, spins are oscillated with the effective staggered fields
  (Fig.~\ref{fig:one-magnon} (c)),
  which induces one-magnon resonance.
  On the other hands, for $E \| b$, 
  $\vec{E}^{\omega} \cdot \vec{\Pi}$ is staggered
  (uniform) along $a$ ($b$) direction.
  Staggered $\vec{E}^{\omega} \cdot \vec{\Pi}$ along $a$ 
  direction does not couple  with the uniform spins along $a$ direction,
  {\it i.e.}, the effective fields cancels out,
  and, thus, no one-magnon resonance occurs.
  
\begin{figure}
    \begin{center}
      \includegraphics[width=0.8\columnwidth]{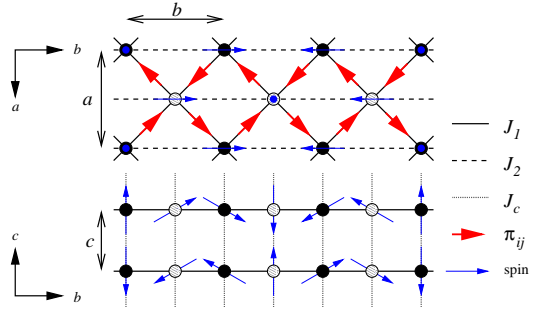}
    \end{center}
    \caption{(Color online) 
      The Heisenberg model for $R$MnO$_3$
      with nearest neighbor interactions $J_1$,
      antiferromagnetic interactions for 2nd neighboring 
      $b$ direction $J_2 (>0)$, and
      antiferromagnetic inter layer interactions $J_c (>0)$.
      The direction of $\vec{\Pi}_{ij}$
      are shown by thick (red) arrows. 
      The direction of $\vec{\Pi}_{ij}$.
      The thin (blue) arrows
      describe the ground state spin configuration for the
      $bc$ cycloidal state.
    }
    \label{fig:lattice}
  \end{figure}

  To proceed more quantitative argument in $R{\rm MnO_3}$, 
  let us consider the frustrated 3-dimensional 
  $S = 2$ Heisenberg Hamiltonian as shown in 
  Fig.~\ref{fig:lattice}
  as the simplest model to reproduce its magnetic behaviors.
  $J_2$ and $J_c$ are fixed to be antiferromagnetic 
  interactions. Although $J_1$ is a ferromagnetic interaction
  in $R{\rm MnO_3}$, we also extended our calculation 
  to the antiferromagnetic $J_1$ region.
  In $R{\rm MnO_3}$, interaction $J_2$ strongly 
  depends $R$ ions. Thus, we treat the ratio $J_2/J_1$ 
  as a parameters and fixes the other parameters
  to be $|J_1| = 8$ meV, and $J_c/|J_1| = 1.5$
  which well reproduces the magnetic behaviors of 
  $R{\rm MnO_3}$ including phase diagrams~\cite{mochizuki08,mochizuki08b}.
  Due to the frustration,
  a spiral spin state is a ground state in the
  parameter range $J_2/|J_1| > 0.5$ for classical spins, and 
  a spiral angle $\theta$ is defined as $\cos \theta = - J_1 / 2 J_2$. 
  An anisotropy term $D \sum_i \left( S_i^a \right)^{2}$
  is added to fix the spiral plane to be $bc$ ($D/|J_1| = 0.2$).
  
  At low temperatures, the magnetic behaviors of
  cycloidal ordered magnets can be described by 
  the linear spin wave theory~\cite{cooper62}.
  After rotations of local axis and  Holstein-Primakoff approximations,
  the spin Hamiltonian is written 
  by the spin wave annihilation (creation) operator
  $\alpha^\dagger_k$ ($\alpha_k$)~\cite{mochizuki08b}.
  The spin-dependent polarization is also represented by
  spin wave annihilation and creation operators.
  The polarization (\ref{eq:polarization_S})
  is written as one- and two-magnon processes:
  \begin{eqnarray}
    \vec{P}_{S} & = & 
    i S \sqrt{S N} \sin \theta \,\, \vec{\Pi}^{(1)}(\vec{k}_{2 \pi})  
    (\alpha_{k_{2\pi}}^{\dagger} - \alpha_{k_{2\pi}})
    \nonumber \\
    & & + i S \sin^2 \frac{\theta}{2} \sum_k \vec{\Pi}^{(2)}(\vec{k})  
    \nonumber \\ & & \hspace*{0.5cm}\times
    \left( \alpha_k^{\dagger} \alpha_{-k-k_{2\pi}}^{\dagger}
    - \alpha_k \alpha_{-k-k_{2\pi}} \right). \,\,\,\,\,\,\,\,
    \label{eq:P_S}
  \end{eqnarray}
  Here $\vec{\Pi^{(1)}}$ has only $a$ component 
  (see Fig~\ref{fig:one-magnon} (c) and (d)). 
  One-magnon absorption can be induced
  only for the condition $E \| a$ but 
  two-magnon absorption occurs for both $E \| a$ and $b$.
  In the one-magnon process, magnons at $\vec{k}_{2 \pi} = (2 \pi/a, 0, 0)$ 
  and $(0, 2 \pi/b, 0)$ are induced and
  each of them has the same magnon energy $\omega_{2 \pi}$.
  Details of $\vec{\Pi^{(2)}}$ will be reported elsewhere.

\begin{figure}
    \begin{center}
      \includegraphics[width=0.9\columnwidth]{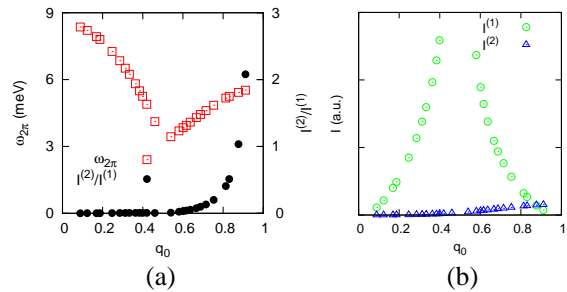}
    \end{center}
    \caption{(Color online) 
      (a) The position of the delta function of a one-magnon absorption
      $\omega_{2 \pi}$ are shown by (red) squares. 
      $|J_1|$ is taken to be $8$ meV.
      The ration of $I^{(2)}$/ $I^{(1)}$ are also plotted 
      by filled (black) circles.
      (b) Intensities of one-magnon absorption $I^{(1)}$ 
      shown by (green) circles and 
      two-magnon absorption $I^{(2)}$ shown by (blue) triangles.
    }
    \label{fig:intensity}
  \end{figure}

  The imaginary part of complex electric polarizability tensor at
  $T = 0$,  which represents an absorption, 
  is obtained from Kubo formula: ${\rm Im} \chi_{\alpha \alpha} (\omega) =
  {\rm Im} \chi^{(1)}_{\alpha \alpha} (\omega)
  + {\rm Im} \chi^{(2)}_{\alpha \alpha} (\omega)$
  where
  \begin{eqnarray}
    {\rm Im} \chi^{(1)}_{\alpha \alpha} (\omega)
    & = & N S^3
    \Pi^2(\vec{k}_{2 \pi}) \sin^2 \theta
    \delta(\omega - \omega_{2\pi}) \delta_{\alpha,\, a}  \\
    {\rm Im} \chi^{(2)}_{\alpha \alpha} (\omega)  
    & =  & S^2 
    \sum_k \Pi^{\alpha}(\vec{k}) \Pi^{\alpha}(\vec{k}) 
    \sin^4 \frac{\theta}{2} \nonumber \\
    & & \hspace*{1.2cm} \times \delta(\omega- \omega_k  - \omega_{-k-2\pi})  
     \label{eq:chi}
  \end{eqnarray} 
  The first term $ {\rm Im} \chi^{(1)}_{\alpha \alpha} (\omega)$ 
  represents the one-magnon absorption.
  The position of the absorption peak
  exists at a band edges at $\vec{k}_{2 \pi}$.
  As a function of $q_0 \equiv \theta / (2 \pi/b)$,
  one-magnon peak positions are shown in Fig.~\ref{fig:intensity} (a).
  The second term corresponds to
  the absorption due to simultaneous two-magnon absorption process. 

  To compare the effects of one- and two-magnon processes
  for $E \| a$,
  integrated intensities defined as
  \begin{eqnarray}
    I^{(\beta)} = \int^\infty_0 
    {\rm Im} \chi^{(\beta)}_{a a} (\omega) d\omega 
    \,\,\,\, (\beta = 1\,\,\,{\rm or}\,\,\,2) 
  \end{eqnarray}
  have been calculated as a function of $q_0$, respectively. 
  The results are shown in Fig.~\ref{fig:intensity} (b).
  Obviously, one-magnon absorption is quite strong 
  in a wide range of parameters except for 
  near ferromagnetic phase $q_0 \sim 0$ and
  antiferromagnetic phase  $q_0 \sim 1$.
  The effects of two magnon absorption are
  negligible except near the N\'{e}el ordered state
  as read off from $I^{(2)}/I^{(1)}$ in
  Fig.~\ref{fig:intensity} (a).
  The behaviors of the two magnon absorption intensity 
  are consistent with the well-known fact
  that it is observed in antiferromagnets
  but not in ferromagnets~\cite{tanabe65,moriya66,moriya68}.
  We conclude that 
  for the ferromagnetic interaction $J_1$ ($q_0 < 0.5$)
  region which $R{\rm MnO_3}$ belongs to, 
  the two-magnon absorption is negligible,
  contrary to our first expectation 
  based on the magnon DOS
  that the absorption originates from 
  the conventional two magnon absorption~\cite{takahashi08}.  
  In a view point of the selection rule, 
  dominancy of this one-magnon excitation is consistent 
  with the experimental results
  since light absorption occurs only for $E \| a$ and
  it is independent of the direction 
  of the cycloidal spin plane. 

\begin{figure}
    \begin{center}
      \includegraphics[width=0.8\columnwidth]{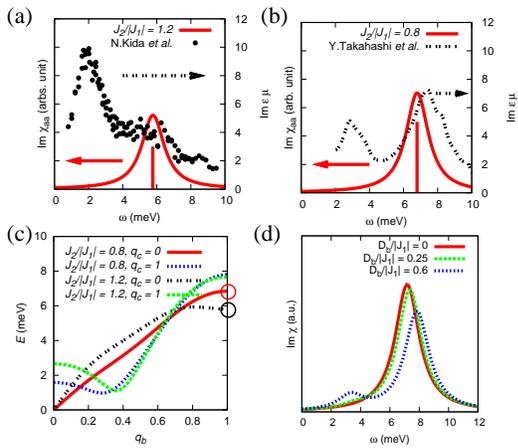}
    \end{center}
    \caption{(Color online) 
      (a) Comparison of spectrum shapes for ${\rm DyMnO_3}$
      between calculation for $J_2/|J_1| = 1.2$
      and experimental data  Im[$\epsilon \mu$] 
      extracted from Ref.~\onlinecite{kida08}.
      (b) Comparison of spectrum shapes for ${\rm TbMnO_3}$
      between calculation for $J_2/|J_1| = 0.8$
      and experimental data  Im[$\epsilon \mu$] 
      extracted from Ref.~\onlinecite{takahashi08}.
      (c) Dispersion relation of the magnon along $b$ direction
      as a function of $q_b$, at $q_a = 0$.
      Circles indicates the magnon energy induced by electric field.
      (c) Spectrum shapes for $J_2/|J_1| = 1/\sqrt{2}$
      with an anisotropy $D_b$.
    }
    \label{fig:spectram}
  \end{figure}

  Let us discuss the spectral shapes.
  As typical examples, we compare 
  ${\rm Im} \chi^{(1)}_{aa}$ with 
  imaginary parts of $\epsilon\mu$ spectra
  in  DyMnO$_3$~\cite{kida08} and 
  TbMnO$_3$~\cite{takahashi08} 
  on the assumption that 
  ${\rm Im} \epsilon\mu \sim \epsilon^{\prime\prime}
  \equiv {\rm Im} \chi^{(1)}_{aa}$. 
  We use interactions $J_2/J_1 = -1.2$
  and $J_2/J_1 = -0.8$ 
  to reproduce cycloidal angles in DyMnO$_3$ ($q_0 \sim 0.36$)
  and TbMnO$_3$($q_0 \sim 0.29$) respectively. 
  Calculated spectra ${\rm Im} \chi^{(1)}_{aa}$ (arb. unit)
  together with experimental data in Refs.~\onlinecite{kida08} 
  and \onlinecite{takahashi08}
  are shown in Figs.~\ref{fig:spectram} (a) and (b). 
  Delta functions are replaced by the Lorentzian with
  width $\epsilon = 1$. 
  The dispersions for magnons are shown 
  in Fig.~\ref{fig:spectram} (c).
  For ${\rm TbMO_3}$, the results are consistent with
  the experimental data in Ref.~\onlinecite{senff07}.
  As shown in Fig.~\ref{fig:spectram} (b),
  our results reproduce the 
  higher energy peak observed in ${\rm TbMnO_3}$ at $\omega \sim 8$ meV.
  In addition, the shoulder structure in ${\rm DyMnO_3}$  
  at $\omega \sim 5$ meV might be interpreted as
  one-magnon absorption as in Fig.~\ref{fig:spectram} (a).
  However, the origin of the lower energy peaks with substantial 
  oscillator strength around $2$ meV is not clear yet.
  More obviously in ${\rm DyMnO_3}$,
  one-magnon absorption mechanism alone can not explain 
  the whole spectra in $R{\rm MnO_3}$. 
 
  Finally, we mention about the effects of
  anisotropy terms. Once the superstructure is induced 
  by an anisotropy, one may think that such low 
  energy peaks can be generated
  because a magnon at zone boundary
  mixes with magnons around zone center. 
  However, satellite peaks induced by superstructures 
  are likely too small to reproduce an experimental 
  observation. For example,
  in $R{\rm MnO_3}$, a spin fan structure is proposed
  by including the effects of anisotropy~\cite{mochizuki08,mochizuki08b},
  which explains well an elliptical modulation
  observed experimentally~\cite{arima06,yamasaki07b}.
  Thus, we examine the effect of easy axis anisotropy 
  along $b$ axis $-\sum D_b (S^b)^2$ for $J_2/|J_1| = -1/\sqrt{2}$,
  ($q_0 = 0.25$), where a fan structure is realized.
  The results are shown in Fig.~\ref{fig:spectram} (d).
  The absorption at low energies appears 
  but the intensity is too small compared with the
  observed peak in ${\rm TbMnO_3}$ even in the case for a rather 
  strong anisotropy, {\it i.e.} $D_b/|J_1| = 0.6$.
  In this way, such an effect produces a smaller satellites
  and fails to explain a larger peak as in ${\rm DyMnO_3}$
  except for the unlikely case that modulations of the cycloidal
  spin structure are extremely large.
  
  So far, it has been believed that the electric component of light 
  can not induce a one-magnon resonance in Heisenberg systems
  through $\vec{S}_i \cdot \vec{S}_j$,
  because the symmetric spin polarization $(\ref{eq:polarization_S})$ 
  conserves the total spin moment
  as seen in a two-magnon absorption 
  in N\'{e}el ordered state~\cite{tanabe65,moriya66,moriya68}.
  However, we can now predict that one-magnon resonances can be induced
  by oscillating electric field
  even in some non-collinear Heisenberg system.
  This becomes possible because the ground state itself
  is symmetry broken, which makes a matrix element
  between ground state and one-magnon excitations non-zero.
  In this paper, we have clarified that one-magnon absorption 
  is possible for a certain type of cycloidal spin states.
  Moreover, one-magnon resonance might also be observed in
  some other non-colinear spin systems, {\it e.g.}
  canted N\'{e}el ordering state
  as we can imagine from Fig.~\ref{fig:one-magnon} (b). 
  Our results indicate the possibility that electromagnons 
  are detectable in a wide range of materials. 

  In terahertz absorption in $R{\rm MnO_3}$, 
  a few mystery are still left as open issues.
  The most challenging and important point is
  to clarify the origin of the lower energy absorption around $2$meV.
  Also in absorption around $5 \sim 8$meV,
  we need a rather large damping factor 
  to reproduce experimental results, but 
  the reason of its largeness is not obvious.
  Such a rather widely spreaded spectra are also observed 
  even in conventional magnon by the inelastic 
  neutron scattering experiments~\cite{senff07},
  thus, that might be come from the peculiar features of
  the magnon in the cycloidal spin state. 
  The $E \| a$ absorption is observed even 
  in the high temperature collinear phase, {\it i.e.}, 
  paraelectric phase~\cite{kida08,kida08b}. 
  Within the linear spin wave theory, 
  we can not treat the collinear phase properly.
  However, if we assume that 
  the collinear phase is realized as a superposition of
  nearly degenerate $ab$ and $bc$ cycloidal planes
  as proposed in Refs.~\onlinecite{mochizuki08}
  and \onlinecite{mochizuki08b},
  the one-magnon absorption due to symmetric term 
  might occur even in the collinear phase within our framework.
  
  We thank H. Katsura and N. Nagaosa for useful comment,
  and N. Kida, M. Mochizuki, Y. Takahashi, T. Arima, R. Shimano,
  and Y. Tokura for fruitful discussion.
  This work is in part supported by Grant-In-Aids for Scientific 
  Research from the Ministry of Education, Culture, 
  Sports, Science and Technology (MEXT) Japan.
  
  {\it Note added}---Recently, we became aware of 
  similar work by R. Vald\'{e}s Aguilar~\cite{aguilar08},
  where they report on absorption in ${\rm TbMnO_3}$
  and indicate the existence of one-magnon absorption
  considering the effects of orthorhombic lattice distortions.
 
  \bibliographystyle{apsrev}
  \bibliography{EM}

\end{document}